\documentclass[12pt]{article}
\usepackage{epsfig}

\usepackage{color}
\usepackage{psfrag}
\usepackage{amsmath,amssymb}

\newcommand{\be}{\begin{equation}}
\newcommand{\ee}{\end{equation}}

\newcommand{\bea}{\begin{eqnarray}}
\newcommand{\eea}{\end{eqnarray}}

\newcommand{\beas}{\begin{eqnarray*}}
\newcommand{\eeas}{\end{eqnarray*}}

\newcommand{\nn}{\nonumber}

\newcommand{\bd}{\begin{displaymath}}
\newcommand{\ed}{\end{displaymath}}

\newcommand{\bei}{\begin{itemize}}
\newcommand{\eei}{\end{itemize}}

\newcommand{\IM}{{\rm Im}}

\newcommand{\Br}{{\rm Br}}

\newcommand{\rC}{r_{\textrm{C}}}

\newcommand{\rEW}{r_{\textrm{EW}}}

\newcommand{\rtEW}{\tilde{r}_{\textrm{EW}}}

\newcommand{\rtEWA}{\tilde{r}_{\textrm{EW}}^{\textrm{A}}}

\def\beq{\begin{equation}}
\def\eeq#1{\label{#1}\end{equation}}
\def\eeqn{\end{equation}}

\def\beqa{\begin{eqnarray}}
\def\eeqa#1{\label{#1}\end{eqnarray}}
\def\eeqan{\end{eqnarray}}

\let\bar=\overbar

\def\Dslash{\not{\hbox{\kern-4pt $D$}}}
\def\dslash{\not{\hbox{\kern-2pt $\del$}}}

\def\ee{e^+e^-}

\def\msb{{\bar{\ssstyle M \kern -1pt S}}}

\def\Title#1{\begin{center} {\Large {\bf #1} } \end{center}}

\begin{document}

\begin{flushright}
Proceedings of CKM 2012, the 7th
International Workshop on the CKM Unitarity \\
Triangle,  University of Cincinnati,
USA, 28 September - 2 October 2012.
\end{flushright}

\Title{\boldmath
Status of the $B\to \pi K$ puzzle and its relation\\
to $B_s\to\phi\pi$ and $B_s\to\phi\rho$ decays }

\bigskip\bigskip

%+\addtocontents{toc}{{\it L. Hofer, L. Vernazza}}
%+\label{VernazzaStart}

\begin{raggedright}

{\it Lars Hofer$^{a}$ and Leonardo Vernazza$^{b,c,}$\footnote{Speaker}\index{L. Hofer, L.Vernazza}\\[0.2cm]
(a) Institut f\"ur Theoretische Physik und Astrophysik, Universit\"at W\"urzburg, \\
D-97074 W\"urzburg, Germany \\[0.2cm]
(b) PRISMA Cluster of Excellence \& Institut f\"ur Physik (THEP) \\
Johannes Gutenberg-Universit\"at, D--55099 Mainz, Germany  \\[0.2cm]
(c) Dipartimento di Fisica, Universit\`a di Torino, \& INFN, Sezione di Torino, \\
Via P. Giuria 1, I-10125 Torino, Italy \\[0.2cm]}

\bigskip\bigskip
\end{raggedright}

\begin{abstract}
\noindent Some discrepancies between theory and experiment
in the $B\to\pi K$ decays suggest the possibility of a new
physics contribution with the structure of an electroweak
penguin amplitude. If such a scenario is realised in nature,
the branching ratios of the decays $B_s\to \phi \pi$ and
$B_s\to\phi\rho$ can be enhanced by about one order of magnitude.
We review and update the current status of the $B\to\pi K$ puzzle
and its implications for the decays $B_s\to \phi\pi$ and
$B_s\to\phi\rho$.
\end{abstract}

\section{Status of the $B\to \pi K$ puzzle}

The  $B\to\pi K$ decays provide a useful test of the flavour structure
and of CP violation in the Standard Model (SM). They have a small
branching ratio, $\mathcal O(10^{-6})$, and are sensitive to
contributions from New Physics (NP). In the last decade some discrepancies between
$B\to\pi K$ measurements and SM predictions have occurred, leading to
speculations of a ``$B\to\pi K$ puzzle'' \cite{BpiKpuzzle}.
The measured values of the branching fractions and CP asymmetries have
fluctuated around the SM predictions, giving in some cases
discrepancies up to $3 \sigma$ between theory and experiment,
but none of them was large enough and stable over a longer period to provide a clear
indication of NP. In Fig. \ref{fig1} we give an updated comparison
of experimental and theoretical results for the $B\to\pi K$ observables $R_i$ and
$\Delta A_{\rm CP}$, which represent ratios of branching fractions and
differences of direct CP asymmetries,
as defined in \cite{HoferSchererVernazza}. The results have been
obtained employing QCD factorisation (QCDF) \cite{QCDF} for the evaluation of hadronic matrix elements
and using the most updated experimental data and theoretical
input \cite{Amhis:2012bh}. The observable which shows
the largest discrepancy ($\sim2.2\sigma$) between theory and experiment is
\begin{equation}\label{a1}
\Delta A_{\rm CP} \equiv A_{\rm CP}(B^-\to\pi^0 K^-)
-A_{\rm CP}(\bar{B}^0\to\pi^+ K^-)
\stackrel{\text{SM}}{=} 1.9^{+5.8}_{-4.8}\,\%
\stackrel{\text{exp.}}{=} (12.6 \pm 2.2)\, \%.
\end{equation}
Referring to \cite{HoferSchererVernazza}
for a detailed analysis, we summarise the status of the
``$B\to\pi K$ puzzle'' as follows:

\begin{figure}[t]
\begin{center}
\epsfig{file=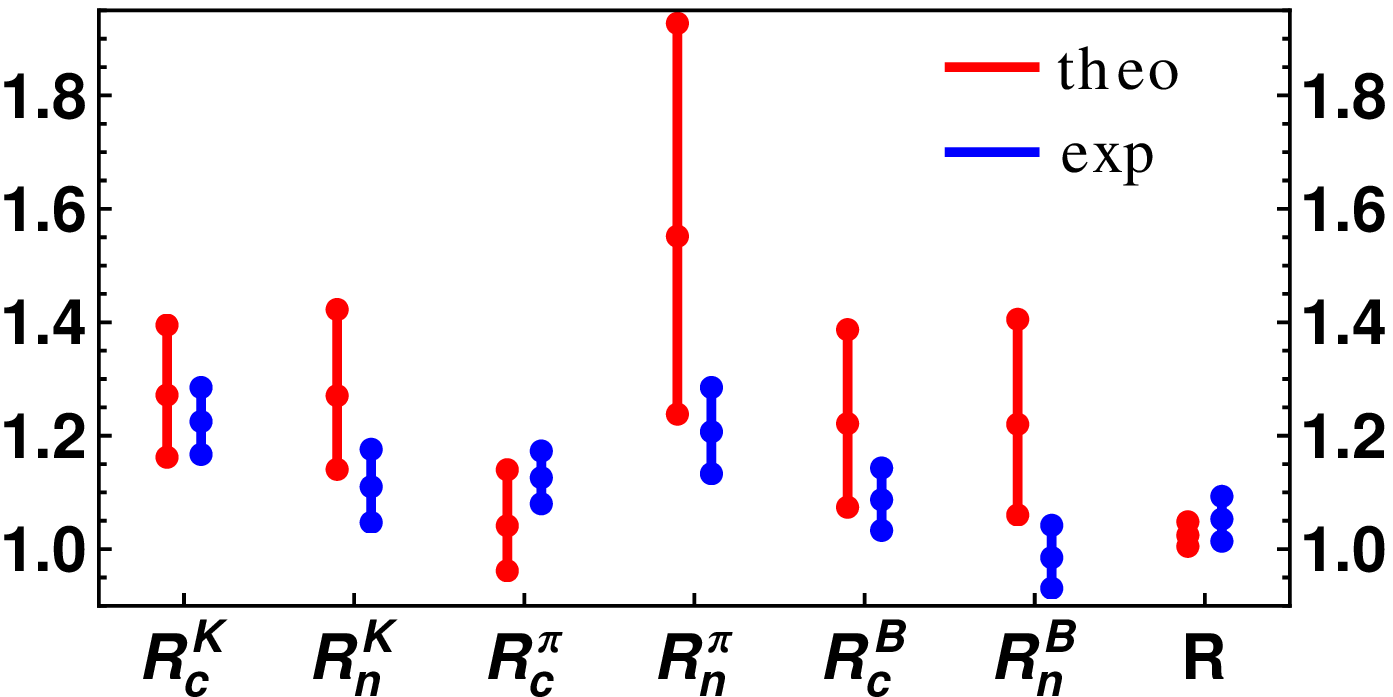,height=3.48cm} \hspace{0.2cm}
\epsfig{file=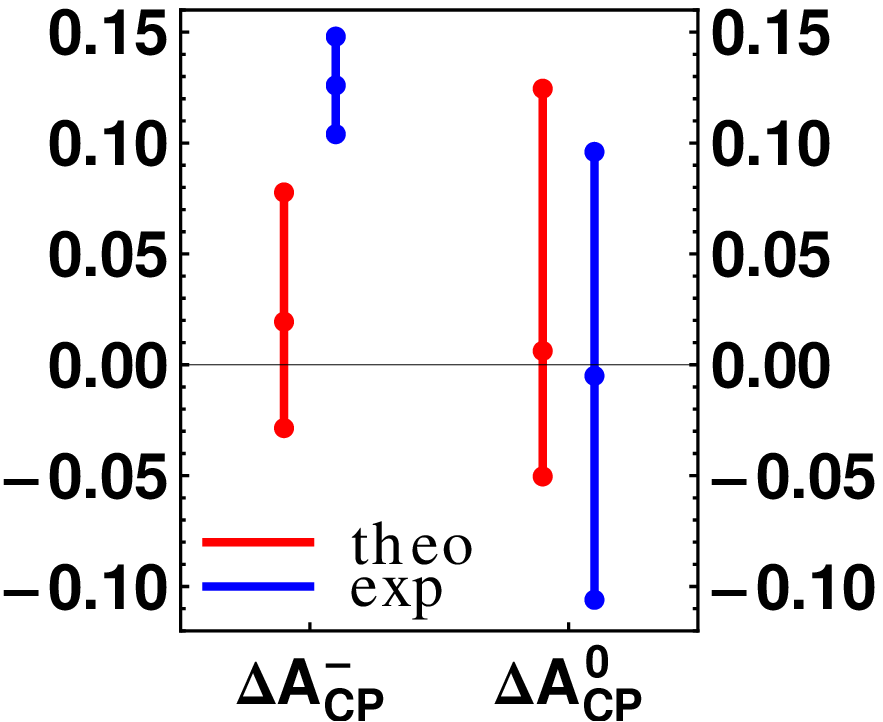,height=3.48cm} \hspace{0.2cm}
\epsfig{file=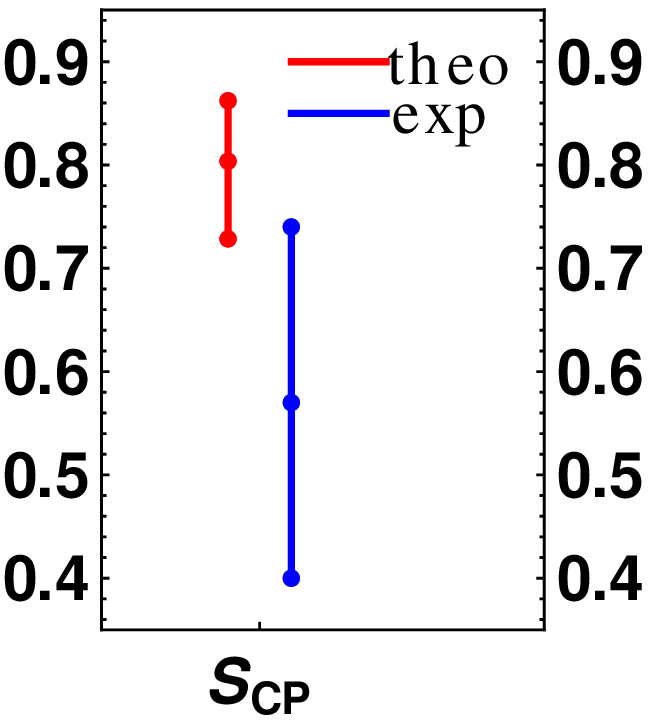,height=3.48cm}
\caption{\footnotesize Comparison of theoretical preditions and experimental
results for various observables for the $B\to\pi K$ decays,
as defined in \cite{HoferSchererVernazza}.}
\label{fig1}
\end{center}
\end{figure}

\bei
\item Direct CP asymmetries arise through the interference
of weak and strong phases. They are small ($\leq 10\%$) in the framework of QCDF
because strong phases are
either perturbative and ${\cal O}(\alpha_s)$-suppressed,
or non-perturbative and
${\cal O}(\Lambda_{\rm QCD})$-suppressed. Therefore it is difficult
to explain the experimental result (\ref{a1}) in
the SM.
\item In the SM, $\Delta A_{\rm CP}$ can be topologically parametrised as
\begin{equation}\label{g10}
\Delta A_{\rm CP}\,\simeq\,-2\,\IM \left(r_{\rm C}\right) \sin\gamma,
\end{equation}
where $r_{\rm C}$ is the colour-suppressed tree-level amplitude, normalised to the dominant
QCD penguin amplitude, and $\gamma$ is the weak CKM phase.
A large $\Delta A_{\rm CP}$ can be explained
only with an enhanced color-suppressed tree amplitude,
in contrast with QCDF expectations. Adding, on the other hand, a new
isospin-violating amplitude (again normalised to the QCD penguin amplitude)
\begin{equation}\label{g8}
r_\text{EW}  \rightarrow r_\text{EW} + \tilde{r}_\text{EW}
e^{-i\delta},\hspace{0.6cm}
r_\text{EW}^{\textrm{C}}  \rightarrow r_\text{EW}^{\rm C} +
\tilde{r}_\text{EW}^{\rm C} e^{-i\delta},
\hspace{0.6cm}
r_\text{EW}^{\textrm{A}}  \rightarrow r_\text{EW}^{\rm A} +
\tilde{r}_\text{EW}^{\rm A} e^{-i\delta},
\end{equation}
 with a new weak phase $\delta$ and with the structure of an EW penguin
topology ($r_{\rm EW}$: colour-allowed, $r_{\rm EW}^{\rm C}$: colour-suppressed,
$r_{\rm EW}^{\rm A}$: annihilation), leads to
\begin{equation}\label{g11}
\Delta A_{\rm CP}\,\simeq\,-2\,\IM \left(r_{\rm C}\right) \sin\gamma +
2\,\IM\left(\tilde{r}_\text{EW} +
\tilde{r}_\text{EW}^\textrm{A}\right)\sin\delta.
\end{equation}
As a consequence, $\Delta A_{\rm CP}$ can turn out much larger in such a scenario than in the SM.
\item Theory uncertainties do not
allow to extract more information from
$B\to\pi K$ decays alone: The EW penguin and the colour-suppressed
tree amplitude always come together in the combination
\begin{equation}\label{eq:ParaCombi}
    \rEW\,-\,\rC\,e^{-i\gamma},
\end{equation}
and $\Delta A_{\rm CP}$, which is the most sensitive
observable to this combination, is proportional to
its imaginary part, which suffers from the largest
uncertainties in QCDF. Other observables, such as
the ratios $R_c^B$ and $R_n^K$ or as
\bea\label{other}\nn
&&\hspace{-1.2cm} S_{\rm CP}(\bar{B}^0\to \pi^0 \bar{K}^0)  \simeq   \sin 2 \beta
        + 2 \mbox{Re} \left( r_{\rm C}\right) \cos 2 \beta \sin\gamma
         -2 \mbox{Re}(\tilde r_{\rm EW}+\tilde r_{\rm EW}^C) \cos 2\beta \sin\delta \\
&&\hspace{1.5cm} \stackrel{\text{SM,QCDF}}{=} \hspace{0.5cm}0.80^{+0.06}_{-0.08}
         \hspace{0.5cm} \stackrel{\text{exp}}{=}\hspace{0.5cm}0.57^{+0.17}_{-0.17},
\eea
involve the real part, whose evaluation
is on a firmer ground. Still,
the EW penguin contribution cannot be disentangled from the dominant QCD penguin and its
uncertainties in any of these observables,
as they are generated via interference of the two contributions.
As an aside, we show in (\ref{other}) the theoretical vs.
experimental prediction for the time-dependent CP asymmetry
$S_{\rm CP}$, and we note that the current value can be
accommodated more easily within the NP hypothesis.
\eei
In order to gain more insight into the problem,
our proposal \cite{HoferSchererVernazza} is
to exploit the large variety of non-leptonic $B$
decays into two charmless mesons, with a look in particular
at the isospin-violating decays
$\bar B_s\rightarrow \phi \rho^0$ and
$\bar B_s\rightarrow \phi \pi^0$, which are dominated
by EW penguins, as pointed out for the first time in
\cite{Fleischer}. We have shown that if NP in this
sector exists at a level where it can explain the
$\Delta A_{\rm CP}$ puzzle, it could be visible in
these purely isospin-violating decays. We update
here the results with the latest experimental and
theoretical input, referring to
\cite{HoferSchererVernazza} for a more
detailed discussion.

\section{The decays $B_s\to\phi\pi$ and $B_s\to\phi\rho$}

The decays $B_s\to\phi\pi$ and $B_s\to\phi\rho$ are purely isospin-violating
and their topological structure is very simple.
Factorising the dominant EW penguin amplitude, the
decay amplitude reads
 \begin{equation} \label{g13}
   \sqrt{2}\,A(\bar{B}_s\to\phi
M_2)\,=\,P_{\textrm{EW}}^{M_2}\,\left(1\,-\,r^{M_2}_{\textrm{C}}\,e^{-i\gamma}\,
+\,\tilde{r}_{\textrm{EW}}^{M_2}\,e^{-i\delta}\right)\,,
 \end{equation}
where $M_2 = \pi$ or $\rho$, $r_{\textrm{C}}^{M_2}$ is the
ratio of the color-suppressed tree to the EW penguin amplitude,
and $\tilde{r}_{\textrm{EW}}^{M_2}$ represents a new EW penguin amplitude
with a new weak phase $\delta$. Even though we are facing the same type
of EW penguin vs colour-suppressed tree amplitude pollution as
in the $B\to\pi K$ decays, the analysis
of $B_s\to\phi\pi$ and $B_s\to\phi\rho$ may be interesting:
\begin{itemize}
 \item A NP amplitude of the order of magnitude needed
to solve the $\Delta A_{\rm CP}$ problem, i.e. of the same order
of the SM EW penguin amplitude, can easily enhance the decays
$B_s\to\phi\pi$ and $B_s\to\phi\rho$ up to one order of magnitude.
 \item The branching ratio is proportional to the real part of
 $r_{\textrm{C}}^{M_2}$ and $\tilde{r}_{\textrm{EW}}^{M_2}$, whose
 evaluation within QCDF is more reliable compared to its imaginary
 part; moreover, compared to $B\to\pi K$ observables
we avoid additional uncertainties coming from the interference with a QCD penguin amplitude,
 which is absent in $B_s\to\phi\pi$ and $B_s\to\phi\rho$.
 \item The helicity structure of these $B_s$ decays is different from the one of
$B\to \pi K$ decays, as the former are $PV$, $VV$ decays, respectively,
while the latter are $PP$ decays, with $P$ denoting a pseudoscalar and $V$ denoting a vector meson.
The non-perturbative low-energy QCD
dynamics is expected to be not correlated among the different classes of
decays, and the exact relation cannot be determined within QCDF.
A NP contribution is instead of high-energy origin, and its effects can
be reliably studied within perturbation theory, allowing for a correlation
among the $B\to \pi K$ and the $B_s$ decays of interest.
\end{itemize}

\section{Model independent analysis}

\begin{figure}[t]
\begin{center}
  \includegraphics[width=0.75\textwidth]{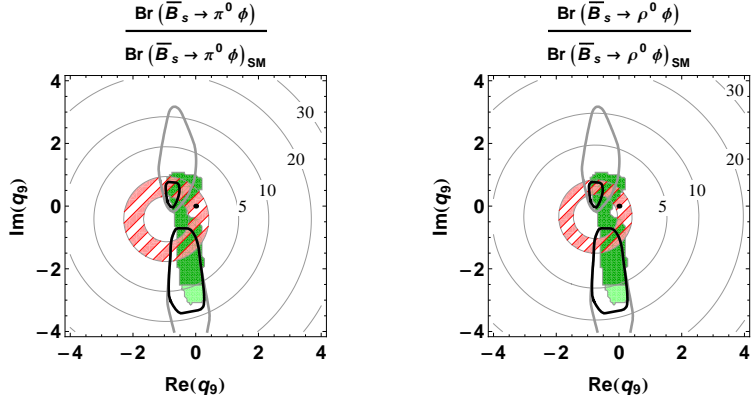}
  \caption{\footnotesize Enhancement factors of the $\bar{B_s}\rightarrow~\phi\rho^0,\phi\pi^0$
  branching ratios with respect to their SM values. The black dot represents the
  SM result while the red striped region shows the theoretical uncertainty
  in the SM. The dark green area is the region allowed by the $2\,\sigma$
  constraints from $\bar B \to \pi K^{(*)},\rho K^{(*)},\phi K^{(*)}$
  and $\bar B_s\to\phi\phi,\bar K K$ decays; for comparison, the light green
  area represents the region allowed by constraints from isospin-sensitive
  observables only, i.e. the $\bar B \to \pi K,\pi K^{(*)},\rho K$ decays.
  The solid black line represents the $1\sigma$ CL of the fit with
  $S_{CP}(\bar B^0 \to \pi \bar{K}^0)$, while the solid grey line
  represents the $1\sigma$ CL of the fit without it.
  Here we show the scenario $q_9\neq 0$.}
  \label{figq9}
\end{center}
\end{figure}

\begin{figure}[t]
\begin{center}
  \includegraphics[width=0.75\textwidth]{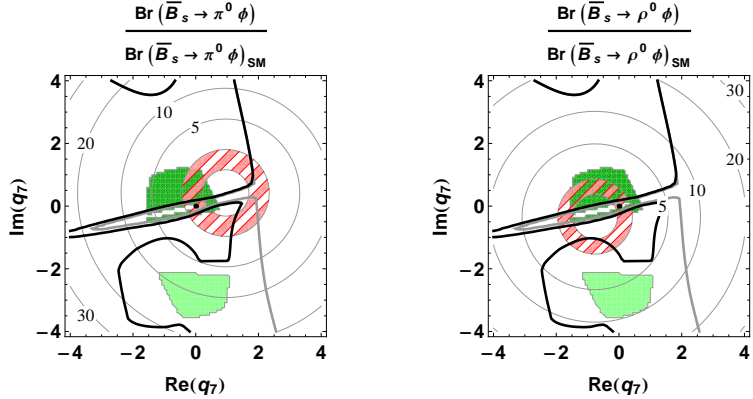}
  \caption{\footnotesize Enhancement factors of the $\bar{B_s}\rightarrow~\phi\rho^0,\phi\pi^0$
  branching ratios with respect to their SM values. The meaning of the contours and regions
  is the same as in fig.~\ref{figq9}. Here we display the scenario $q_7\neq0$.
  The $1\sigma$ region of the fit is the region to the right of the black (grey) curve.}
  \label{figq7}
\end{center}
\end{figure}

We consider the EW penguin operators
\begin{eqnarray}\nn
        Q_7 = (\bar s_{\alpha} b_{\alpha})_{V-A}
        \textstyle{ \sum_q\frac{3}{2}}e_q (\bar q_{\beta}
        q_{\beta})_{V+A}, \quad && Q_8
        = (\bar s_{\alpha} b_{\beta})_{V-A}
        \textstyle{ \sum_q\frac{3}{2}}e_q (\bar q_{\beta} q_{\alpha})_{V+A},\\
        Q_9 = (\bar s_{\alpha} b_{\alpha})_{V-A}
        \textstyle{ \sum_q\frac{3}{2}}e_q (\bar q_{\beta} q_{\beta})_{V-A},
        \quad && Q_{10}
        = (\bar s_{\alpha} b_{\beta})_{V-A}
        \textstyle{ \sum_q\frac{3}{2}}e_q (\bar q_{\beta} q_{\alpha})_{V-A},
\end{eqnarray}
and we parameterise a NP contribution to the
Wilson coefficients as
\begin{equation}\label{g23}
C_{7,9}^{(\prime)\rm NP}(m_W) = C_9^{\rm LO}(m_W)\,q_{7,9}^{(\prime)},
\qquad \qquad q_{7,9}^{(\prime)}  = |q_{7,9}^{(\prime)}|e^{i
\phi_{7,9}^{(\prime)}}.
\end{equation}
Here $C_9^{\rm LO}$ denotes the leading-order SM coefficient and the primed operators are
obtained from the SM $Q_{7,9}$ by flipping the chiralities of the quark fields.
We perform a $\chi^2$-fit to determine the NP parameters
which best describe the $B\to \pi K$ data. Further
hadronic decays like $B\to \rho K,\pi K^*,\rho K^*$ are used to impose
additional constraints at the $2\,\sigma$ level. The result of the fit is
used to study the decays $\bar B_s\to\phi\pi^0,\phi\rho^0$ and to quantify a
potential enhancement of their branching fractions. Such an analysis,
correlating different hadronic decay modes, is only possible if hadronic
matrix elements are calculated from first principles like in
QCDF because the decays $\bar B_s\to\phi\pi^0,\phi\rho^0$
are not related to any other decay via $SU(3)_{F}$ symmetries.

From eq.~(\ref{g11}) we see that the $\Delta A_{\textrm{CP}}$ discrepancy can be
solved either through $\rtEW$ or through $\rtEWA$, which are
both induced by the NP coefficients $q_{7,9}^{(\prime)}$.
Except for parity-symmetric models (where contributions from primed and unprimed
operators exactly cancel), any scenario with at least one of the $q_{7,9}^{(\prime)}$ different from zero
can achieve such a solution.
The minimal $|q|$\,-\,value needed to reduce the $\Delta A_{\textrm{CP}}$
tension below the $1\,\sigma$ level varies among different scenarios. One finds
e.g. $|q_7|\gtrsim 0.3$ for a model with only $q_7\not=0$, $|q_9|\gtrsim 0.8$
for a model with only $q_9\not=0$ and $|q_7|=|q_9|\gtrsim 0.4$ for a model
with $q_7=q_9$ and the primed coefficients being zero.  We note however that the solution of the
$\Delta A_{\textrm{CP}}$ discrepancy via a minimal $|q|$\,-\,value requires
a tuning of the phases $\phi_i$. Realistic scenarios have larger $|q|\sim 1$ values.

It turns out that the annihilation coefficient
$\tilde{r}_{\textrm{EW},\,7^{(\prime)}}^{\textrm{A}}$ induced by $q_7^{(\prime)}$
develops a large imaginary part, so that in scenarios with non-vanishing $q_7^{(\prime)}$
this term gives the dominant contribution to $\Delta A_{\textrm{CP}}$. This explains why in the
$q_7=q_9$ case only a small NP contribution is needed, even though $\rtEW$ is quite small in this scenario, and it
demonstrates the importance of the annihilation
term $\rtEWA$.

For the $q_7$-only and the $q_9$-only scenarios, the result
of the fit to $B\to\pi K$ data and the consequences for the
$B_s$ decays of interest are shown in Figs.~\ref{figq9}, \ref{figq7}.
For details about the fit and the observables used we refer to
\cite{HoferSchererVernazza}. We find  that the $B\to\pi K$ and related
decays set quite strong constraints on the parameter space.
Compared to \cite{HoferSchererVernazza}, the most significant
update here is the inclusion of the constraints from the full set
of branching ratios and CP asymmetries from the $B\to\rho K^*$
decays, which were not available two years ago. We find that,
with inclusion of these decays, the scenarios with
$q_7^{(\prime)}$ are now better constrained. Values $|q_i|\gtrsim 5$
($|q_i|\gtrsim 10$ if only observables sensitive to isospin-violation
are taken into account) are ruled out, i.e. NP corrections cannot be much larger
than the EW penguins of the SM. The SM point is always
excluded at the $2\,\sigma$ level as a direct consequence of the
$\Delta A_{\rm CP}$ data.

From Figs.~\ref{figq9}, \ref{figq7}
the enhancement $\Br^{\textrm{SM}+\textrm{NP}}/\Br^{\textrm{SM}}$ of the
$B_s$ branching fractions can be read off with respect to the different
constraint- and fit-regions. The parts of the allowed region which do
not overlap with the SM uncertainty region are those, where one
expects to be able to distinguish a NP signal without ambiguity. We find
that constraints from other hadronic decays allow in general an enhancement of
$\bar B_s\to\phi\pi^0,\phi\rho^0$ by a factor of $5$ compared to the SM expactation.

In summary, we see that a measurement of $\bar{B}_s\to\phi\pi,\phi\rho$
would complement the data from $B\to \pi K$.
On the one hand, one would be able to test the $\Delta A_{\textrm{CP}}$
data through an enhancement of the $B_s$ decays, as expected
for the $q_7^{(\prime)}\neq 0$, $q_9^{(\prime)}\neq 0$ scenarios;
on the other, we stress that the $B_s$ decays would be
useful in order to distinguish among opposite-parity scenarios,
which cannot be done with the $B\to \pi K$ data alone. An analysis
of $B\to \pi K$ should be supported by the analysis of $PV$
decays, suggesting $\bar B_s\to\phi\pi^0$ as an ideal candidate.

\section{Analysis of specific NP models}

It is opportune to confront the results
obtained with the model independent analysis
with the additional constraints which arise when
considering concrete NP models. In particular,
one has to deal with additional correlations
with the semileptonic decay $\bar{B}\to X_se^+e^-$,
the radiative decay $\bar B\to X_s \gamma$ and
$B_s$-$\bar{B}_s$ mixing. In
\cite{HoferSchererVernazza} we considered three
quite general models, a \emph{modified $Z^0$-penguin scenario},
a \emph{model with an additional $U(1)'$ gauge symmetry}
and the \emph{MSSM}, concluding that the processes above
set quite strong constraints and do not allow
significant enhancement of the decays
$\bar{B}_s\to\phi\pi,\phi\rho$.
Given this result, an updated analysis
with the data from the past two years
does not change the conclusions.
The most significant update
concerns $B_s$-$\bar{B}_s$ mixing,
\cite{Lenz:2012az}, which now points
in the direction of the SM,
in contrast to the $3.8\sigma$ discrepancy
at the time of \cite{HoferSchererVernazza}.
As a consequence,
$B_s$-$\bar{B}_s$ mixing now supports
the results obtained by means of the
semileptonic $B$ decay, which gives the
strongest constraint and allows
at most a factor $\leq 3$ enhancement
of the decays $\bar{B}_s\to\phi\pi,\phi\rho$
decays. To give an idea, we provide
a plot of the allowed enhancement
for the $B_s$ decays in the
modified $Z^0$-penguin scenario
with a non-standard right-handed
$s\bar b Z^0$ coupling.

\begin{figure}[t]
\begin{center}
  \includegraphics[width=0.78\textwidth]{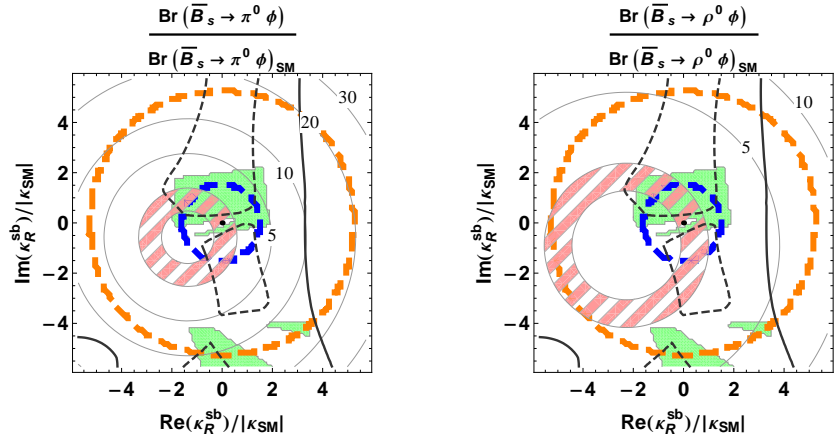}
\end{center}
  \caption{\footnotesize Enhancement factor for the $\bar{B_s}\rightarrow~\phi\rho^0,\phi\pi^0$
  branching ratios with respect to their SM values in a modified-$Z^0$-penguin
  scenario with right-handed coupling. The green area represents the region allowed by the $2\sigma$
constraints from all the considered hadronic decays, while the areas inside the
dashed blue line and inside the dashed orange line represent the regions allowed by the $2\sigma$ constraint
from semi-leptonic decays and from $B_s$-$\bar{B}_s$ mixing, respectively.}
  \label{figZ0}
\end{figure}

\section{Conclusion}

We have updated our work \cite{HoferSchererVernazza},
where we analysed the possibility of probing the $B\to \pi K$
``puzzle'' considering the $\bar B_s\to\phi\pi^0,\phi\rho^0$
decays. The conclusion of
\cite{HoferSchererVernazza} is still valid:
a solution of the $B\to \pi K$ discrepancy
can be obtained adding a NP contribution to the EW penguin
operators, of the same order of magnitude as the
leading SM coefficient $C_9^{\textrm{SM}}$, which in turn
gives rise to an enhancement of $\bar B_s\to\phi\pi^0,\phi\rho^0$.
Now we get stronger constraints to the possible enhancement
from the complete experimental data set of the $B\to\rho K^*$
decays, and, if a correlation is assumed in a concrete
model of NP, from the updated results for
$B_s$-$\bar{B}_s$ mixing.
A model independent analysis, where constraints from
other non-leptonic $B$ decays are considered, shows that
an enhancement of the $\bar B_s\to\phi\pi^0,\phi\rho^0$
decays up to $\sim 5$ times the SM branching ratio
is still possible. In a model dependent analysis,
where one needs to take into account constraints
from semileptonic $B$ decay and $B_s$-$\bar{B}_s$ mixing
as well, the possible enhancement is reduced to
be up to $\sim 3$ times the SM value.
An observation of enhanced $\bar B_s\to\phi\pi^0,\phi\rho^0$
branching ratios would be an interesting complement
to the $B\to \pi K$ data, supporting
the picture of NP in the EW penguins.

\bigskip
L. V. thanks the organizers for a very nice conference experience,
the Alexander-von-Humboldt Foundation and the DFG 
(Deutsche Forschungsgemeinschaft, Project ``Anwendung effektiver 
Feldtheorien in der Colliderphysik: Faktorisierung und Resummation 
für Jetprozesse am LHC'') for support. The work of L.H. has been 
supported by the Helmholtz Alliance ``Physics at the Terascale''.

\end{document}